\documentclass[12pt]{amsart}

\newcommand{\R}{\mathbf R}
\newcommand{\x}{\mathbf x}

\newcommand{\e}{\varepsilon}
\newcommand{\vol}{\mathbf{vol}}
\newcommand{\n}{\mathbf n}
\newcommand{\jj}{\mathbf j}

\newcommand{\DD}{\mathcal D}
\newcommand{\CC}{\mathcal C}

\newcommand{\HH}{\mathcal H}
\newcommand{\F}{\mathcal F}

\newcommand{\eqdef}{\stackrel{\rm def}{=}}

\newcommand{\g}{\mathfrak g}

\begin{document}
\title{A mathematical approach to quantum field theory}
\author{Alexander Roi Stoyanovsky}
\begin{abstract}
We develop a mathematical theory of quantization of multidimensional variational principles, and compare it with traditional constructions of quantum field theory.
We conjecture that mathematical realization of quantum field theory axioms, in general, does not exist.
\end{abstract}
\maketitle

\section*{Introduction}

The purpose of this paper is to present an attempt to construct a mathematical version of quantum field theory, which we consider as part of a mathematical generalization of 
the theory of linear partial differential equations to the case when the bicharacteristics are not curves but are (multidimensional) surfaces. We think that this theory can be interesting from
mathematical point of view. We also compare it with traditional quantum field theory. The result is a conjecture that quantum field theory axioms, in general, can be neither
mathematically implemented nor disproved, like Continuum Hypothesis.

The paper consists of three Sections. In \S1 we state equations of classical field theory given by an arbitrary variational principle. In \S2, we construct 
mathematical quantization of these equations. It uses the solution of the 
deformation quantization problem for the Poisson algebra of classical field theory Hamiltonians.
Using noncommutative multiplication of field theory Hamiltonians, we construct the analog of the Schr\"odinger equation for multidimensional variational principles. We call it by the 
{\it renormalized Tomonaga--Schwinger equation}. In \S3,  based on the solutions of this equation, we construct mathematical expressions for the Bogoliubov $S$-matrix, the Wightman functions, and the Green functions
for the $\varphi^d$ model of scalar field in the spacetime $\R^{n+1}$ for arbitrary $d$ and $n$. It seems clear that, if these objects are well defined, i.~e. do not depend on regularization of the theory, then
they satisfy the axioms and the perturbative expansions of quantum field theory. However, these objects are, in general, ill defined. We think that this is the problem not of our approach but of traditional quantum field theory.
We conjecture that quantum field theory is not a mathematical theory, i.~e. its axioms cannot be, in general, mathematically realized.

\section{Equations of classical field theory}

The formalism of classical field theory equations that we shall need has been developed in the papers [1--4] and in the book [5]. For the sake of completeness, let us briefly recall it here.

 Consider the action functional of the form
\begin{equation}
J=\int_\DD F(x^0,\ldots,x^n,\varphi^1,\ldots,\varphi^m,\varphi^1_{x^0},\ldots,
\varphi^m_{x^n})\,dx^0\ldots dx^n,
\end{equation}
where $x=(x^0,\ldots,x^n)$ is a space-time point; $\varphi^i=\varphi^i(x)$ are real smooth field
functions,  $ i=1,\ldots,m$; $\varphi^i_{x^j}=\partial\varphi^i/\partial x^j$,  $ j=0,\ldots,n$, and integration goes over a domain $\DD$ in the space-time $\R^{n+1}$ with the smooth boundary $\CC=\partial\DD$.

\subsection{Formula for variation of action} We shall need the well known formula for variation of action on a domain with moving boundary, which is used, for instance, in derivation of the Noether theorem. Let us recall this
formula. Let $s=(s^1,\ldots,s^n)$ be parameters on the boundary surface $\CC$: $x=x(s)$. Let $\CC(\e): x=x(s,\e)$ be a deformation of the parameterized surface $\CC$ depending on a small parameter $\e$, $x(s,0)=x(s)$,
and let $\varphi^i(x,\e)$ be a deformation of the field function $\varphi^i(x)=\varphi^i(x,0)$ depending on $\e$. Put $\varphi^i(s,\e)=\varphi^i(x(s,\e),\e)$, and let $\delta$ denote the variation, i.~e. the 
differential with respect to the parameter $\e$ at $\e=0$. Let $J=J(\e)$ be the integral (1) with the field functions $\varphi^i(x,\e)$ over the domain $\DD(\e)$ with the boundary $\CC(\e)$. 
Then one has the following formula for variation of the functional $J$:
\begin{equation}
\begin{aligned}{}
\delta J&=\int_\DD\sum\limits_i\left(F_{\varphi^i}-\sum_j\frac{\partial}{\partial x^j}F_{\varphi^i_{x^j}}\right)\delta\varphi^i(x)dx\\
&+\int_\CC\left(\sum\pi_i(s)\delta\varphi^i(s)-\sum H_j(s)\delta x^j(s)\right)ds,
\end{aligned}
\end{equation}
where $F_{\varphi^i}=\partial F/\partial\varphi^i$, $F_{\varphi^i_{x^j}}=\partial F/\partial\varphi^i_{x^j}$,
\begin{equation}
\begin{aligned}{}
\pi_i&=\sum_l F_{\varphi^i_{x^l}}(-1)^l\frac{\partial(x^0,\ldots,\widehat{x^l},\ldots,x^n)}
{\partial(s^1,\ldots,s^n)},\\
H_j&=\sum_{l\ne j}\left(\sum_i F_{
\varphi^i_{x^l}}\varphi^i_{x^j}\right)
(-1)^l\frac{\partial(x^0,\ldots,\widehat{x^l},\ldots,x^n)}
{\partial(s^1,\ldots,s^n)}\\
&+\left(\sum_i F_{\varphi^i_{x^j}}\varphi^i_{x^j}-F\right)
(-1)^j\frac{\partial(x^0,\ldots,\widehat{x^j},\ldots,x^n)}
{\partial(s^1,\ldots,s^n)}.
\end{aligned}
\end{equation}
Here $\frac{\partial(x^1,\ldots,x^n)}{\partial(s^1,\ldots,s^n)}$ is the Jacobian, and the hat over a variable means that the variable is omitted.

If the boundary functions are fixed, $\delta\varphi^i(s)=\delta x^j(s)=0$, then the condition $\delta J=0$ yields the Euler--Lagrange equations
\begin{equation}
F_{\varphi^i}-\sum_j\frac{\partial}{\partial x^j}F_{\varphi^i_{x^j}}=0,\ \ \ i=1,\ldots,m.
\end{equation}
 This is a system of nonlinear second order partial differential equations for the functions $\varphi^i(x)$.

\subsection{The generalized Hamilton--Jacobi equation}
Let us assume that for any functions $x^j(s), \varphi^i(s)$ sufficiently close to certain fixed functions, there exist the unique field functions $\varphi^i(x)$ satisfying the 
Euler--Lagrange equations (4) with the boundary conditions $\varphi^i(x(s))=\varphi^i(s)$. Denote by $S=S(\varphi^i(s),x^j(s))$ the integral (1) with these field functions $\varphi(x)$ over the domain $\DD$ with the boundary $\CC$: $x=x(s)$.
Let us derive the equations satisfied by the functional $S(\varphi^i(s),x^j(s))$.

To this end, note that
by formula (3), the quantities
\begin{equation}
\pi_i(s)=\frac{\delta S}{\delta\varphi^i(s)},\ \ H_j(s)=-\frac{\delta S}{\delta x^j(s)}
\end{equation}
(for the definition of variational derivatives $\delta S/\delta\varphi^i(s)$ and $\delta S/\delta x^j(s)$, see Subsect. 2.0 below) 
depend not only on the functions $\varphi^i(s)$, $x^j(s)$, but also on the derivatives $\varphi^i_{x^j}$. 
These $m(n+1)$ derivatives are related by $mn$ equations
\begin{equation}
\sum_j\varphi^i_{x^j}x^j_{s^k}=\varphi^i_{s^k},\ \ i=1,\ldots,m,\ \ k=1,\ldots,n.
\end{equation}
Therefore, $m+n+1$ quantities (5) depend on $m(n+1)-mn=m$ free parameters. Hence they are related by $n+1$ equations.
$n$~of these equations are easy to find:
\begin{equation}
\sum_i\varphi^i_{s^k}\pi_i-\sum_j x^j_{s^k}H_j=0,\ \ k=1,\ldots,n.
\end{equation}
These equations express the fact that the value of the functional $S$ does not depend on a parameterization
of the surface $\CC$.

The remaining  $(n+1)$-th equation depends on the form of the function $F$. Denote it by
\begin{equation}
\HH\left(x^j(s),\varphi^i(s),x^j_{s^k}(s),\varphi^i_{s^k}(s),\pi_i(s),-H_j(s)\right)=0.
\end{equation}
From equalities (5) we obtain the following system of equations on the functional $S$, called
the {\it generalized Hamilton--Jacobi equation}:
\begin{equation}
\left\{
\begin{array}{l}
\sum\limits_i\varphi^i_{s^k}\frac{\delta S}{\delta\varphi^i(s)}+\sum\limits_j
x^j_{s^k}\frac{\delta S}{\delta x^j(s)}=0,\ \ \ k=1,\ldots,n,\\
\HH\left(x^j(s),\varphi^i(s),x^j_{s^k}(s),\varphi^i_{s^k}(s),
\frac{\delta S}{\delta\varphi^i(s)},\frac{\delta S}{\delta x^j(s)}\right)=0.
\end{array}
\right.
\end{equation}
\medskip

{\bf Examples.}
\nopagebreak
\medskip

1) (Scalar field with self-action) [4, 5] Let $m=1$ and
\begin{equation}
\begin{aligned}{}
F(x^j,\varphi,\varphi_{x^j})&=\frac12\left(\varphi_{x^0}^2-\sum_{j\ne0}\varphi_{x^j}^2\right)-V(x,\varphi)\\
&=\frac12\varphi_{x^\mu}\varphi_{x_\mu}-V(x,\varphi),
\end{aligned}
\end{equation}
where we have introduced Greek indices $\mu$ instead of $j$, and raising and lowering indices goes using the
Lorentz metric
\begin{equation}
dx^2=(dx^0)^2-\sum_{j\ne 0}(dx^j)^2.
\end{equation}
Then the generalized Hamilton--Jacobi equation has the following form:
\begin{equation}
\left\{
\begin{array}{l}
x^\mu_{s^k}\frac{\delta S}{\delta x^\mu(s)}+
\varphi_{s^k}\frac{\delta S}{\delta\varphi(s)}=0,\ \ k=1,\ldots,n,\\
\vol\cdot\frac{\delta S}{\delta\n(s)}+
\frac12\left(\frac{\delta S}{\delta\varphi(s)}\right)^2
+ \vol^2\left(\frac12 d\varphi(s)^2+V(x(s),\varphi(s))\right)=0,
\end{array}
\right.
\end{equation}
where $\vol$ is the volume element on the surface $x=x(s)$,
\begin{equation}
\vol^2=D^\mu D_\mu,\ \  D^\mu=(-1)^\mu \frac{\partial(x^0,\ldots,\widehat{x^\mu},\ldots,x^n)}
{\partial(s^1,\ldots,s^n)},
\end{equation}
$\delta S/\delta\n(s)$ is the normal derivative,
\begin{equation}
\vol\cdot\frac{\delta S}{\delta\n(s)}=D_\mu\frac{\delta S}{\delta x^\mu(s)},
\end{equation}
and $d\varphi(s)^2$ is the scalar square of the differential $d\varphi(s)$ of the function $\varphi(s)$ on the surface $x=x(s)$,
\begin{equation}
\begin{aligned}{}
\vol^2 d\varphi(s)^2=&(D_\mu\varphi_{x^\mu})^2-(D_\mu D^\mu)(\varphi_{x^\nu}\varphi_{x_\nu})=-\sum\limits_{\mu<\nu}\sum\limits_{k,l}D^{\mu\nu}_kD_{\mu\nu,l}\varphi_{s^k}\varphi_{s^l},\\
&D^{\mu\nu}_k=(-1)^{k+\mu+\nu}\frac{\partial(x^0,\ldots,\widehat{x^\mu},\ldots,\widehat{x^\nu},\ldots,x^n)}
{\partial(s^1,\ldots,\widehat{s^k},\ldots,s^n)}.
\end{aligned}
\end{equation}

\medskip

2) (The Plateau problem) [5] Assume that one considers $(n+1)$-dimensional surfaces $D$ with the boundary $C:x=x(s)$
in the Euclidean space $\R^N$
with coordinates $(x^1,\ldots,x^N)$, and the role of integral (1) is played by the area of the surface $D$.
Then the generalized Hamilton--Jacobi equation is
\begin{equation}
\left\{
\begin{array}{l}
\sum\limits_j x^j_{s^k}\frac{\delta S}{\delta x^j(s)}=0,\ \ 1\le k\le n,\\
\sum\limits_j\left(\frac{\delta S}{\delta x^j(s)}\right)^2=
\sum\limits_{j_1<\ldots<j_n}\left(\frac{\partial(x^{j_1},\ldots,x^{j_n})}
{\partial(s^1,\ldots,s^n)}\right)^2.
\end{array}
\right.
\end{equation}

\subsection{The generalized canonical Hamilton equations}

From $n+1$ equations (7, 8) one can, in general, express $H_j$ as functions 
\begin{equation}
H_j=H_j(s)=H_j(x(s),x_{s^k}(s),\varphi^i(s),\varphi^i_{s^k}(s),\pi_i(s)).
\end{equation}
Let $x^j=x^j(s^1,\ldots,s^n,t)$, $j=0,\ldots,n$, be arbitrary functions, and let
\begin{equation}
\begin{aligned}{}
H&=H(t)=H(t,\varphi^i(\cdot),\pi_i(\cdot))\\
&=\int\sum_j H_j(x(s,t),x_{s^k}(s,t),\varphi^i(s),\varphi^i_{s^k}(s),\pi_i(s)) x^j_t(s,t)ds,
\end{aligned}
\end{equation}
where $\varphi^i(\cdot)$, $\pi_i(\cdot)$ are two functions of $s$.
Then the Euler--Lagrange equations (4) are equivalent to the following {\it generalized canonical Ham\-ilt\-on equations}:
\begin{equation}
\left\{\begin{array}{l}
\frac{\partial\varphi^i(s,t)}{\partial t}=\frac{\delta H}{\delta\pi_i(s)}(t,\varphi^i(\cdot,t),\pi_i(\cdot,t)),\\
\frac{\partial\pi_i(s,t)}{\partial t}=-\frac{\delta H}{\delta\varphi^i(s)}(t,\varphi^i(\cdot,t),\pi_i(\cdot,t)).
\end{array}
\right.
\end{equation}

Another form of these equations is the following one. Let 
$$\Phi(x^j(\cdot);\varphi^i(\cdot),\pi_i(\cdot))$$ be a functional of functions $x^j(s)$, $\varphi^i(s)$, $\pi_i(s)$.
Then the condition
\begin{equation}
d\Phi(x^j(\cdot,t);\varphi^i(\cdot,t),\pi_i(\cdot,t))/dt=0
\end{equation}
is equivalent to the equation 
\begin{equation}
\frac{\partial\Phi(x^j(\cdot,t);\varphi^i(\cdot),\pi_i(\cdot))}{\partial t}=\{\Phi(x^j(\cdot,t);\varphi^i(\cdot),\pi_i(\cdot)),H(t)\},
\end{equation}
 where 
\begin{equation}
\{\Phi_1,\Phi_2\}=\int\sum\limits_i\left(\frac{\delta\Phi_1}{\delta\pi_i(s)}\frac{\delta\Phi_2}{\delta\varphi^i(s)}-\frac{\delta\Phi_1}{\delta\varphi^i(s)}\frac{\delta\Phi_2}{\delta\pi_i(s)}\right)ds
\end{equation}
is the Poisson bracket of two functionals $\Phi_1(\varphi^i(\cdot),\pi_i(\cdot))$, $\Phi_2(\varphi^i(\cdot),\pi_i(\cdot))$.

Equivalently, the equation is
\begin{equation}
\frac{\delta\Phi}{\delta x^j(s)}=\{\Phi,H_j(s)\}.
\end{equation}

Let us call by a classical field theory {\it observable} a functional $\Phi(x^j(\cdot)$; $\varphi^i(\cdot)$, $\pi_i(\cdot))$ satisfying equations (20), (21), or (23). Observables form a commutative Poisson algebra with respect to 
multiplication of functionals and the Poisson bracket (22).
Equations (20), (21), or (23) mean that $\Phi$ actually depends not on $x^j(s)$, $\varphi^i(s)$, $\pi_i(s)$ but only on the solution $\varphi^i(x)$ of the Euler--Lagrange equations (4) with the initial conditions 
$\varphi^i(x(s))=\varphi^i(s)$, $\pi_i(x(s))=\pi_i(s)$. 

The system of equations (23) satisfies the Frobenius integrability (zero curvature) condition,
\begin{equation}
\frac{\delta H_j(s)}{\delta x^{j'}(s')}-\frac{\delta H_{j'}(s')}{\delta x^j(s)}-\{H_j(s),H_{j'}(s')\}=0.
\end{equation}
This condition means that the solution of the system (23) is well defined.

\medskip

{\bf Example.} For the scalar field with self-action (10, 12), the system of generalized canonical Hamilton equations is equivalent to the following system of equations for a functional $\Phi(x^\mu(\cdot);\varphi(\cdot),\pi(\cdot))$: 
\begin{equation}
\left\{
\begin{array}{l}
x^\mu_{s^k}(s)\frac{\delta\Phi}{\delta x^\mu(s)}=
\{\Phi,\varphi_{s^k}(s)\pi(s)\},\ \ k=1,\ldots,n,\\
\vol\cdot\frac{\delta\Phi}{\delta\n(s)}=\left\{\Phi,
\frac12\pi(s)^2
+ \vol^2\left(\frac12 d\varphi(s)^2+V(x(s),\varphi(s))\right)\right\}.
\end{array}
\right.
\end{equation}

The Euler--Lagrange equations and the equivalent generalized canonical Hamilton equations are called the characteristics equations for the generalized Hamilton--Jacobi equation. Regarding the integration theory 
of the generalized Hamilton--Jacobi equation using integration of the characteristics equations, see [13] (for the case $m=n=1$) and [2, 5] (for the general case). 

{\bf Problem.} Conversely, integrate Euler--Lagrange equations with large symmetry groups (such as the Einstein or Yang--Mills equations) using integration of the generalized Hamilton--Jacobi equation.

\subsection{Symmetries} Let $G$ be a Lie group of symmetries of the action (1), i.~e. transformations 
$(x,\varphi)\to (\widetilde x,\widetilde\varphi)=P(x,\varphi)$ of variables $(x,\varphi)=(x^j,\varphi^i)$ preserving the integral (1), and let 
$\g$ be the Lie algebra of $G$. Let $\alpha=dP(\e)/d\e|_{\e=0}\in\g$, where $P(\e)$ is a curve in $G$ with $P(0)=1$, and let  
\begin{equation}
\delta\widetilde\varphi^i_\alpha(x,\varphi)=a^i_\alpha(x,\varphi)d\e,\ \  \delta\widetilde x^j_\alpha(x,\varphi)=b^j_\alpha(x,\varphi)d\e
\end{equation}
be the corresponding infinitesimal transformation of the variables $(x^j$, $\varphi^i)$. Let $x^j(s,t)$ be arbitrary functions, let 
$$
\varphi^i(s,t)=\varphi^i(x(s,t)),\ \ \pi_i(s,t)=\pi_i(x(s,t))
$$ 
be the solution of the canonical Hamilton equations (19)
corresponding to a solution $\varphi^i(x)$ of the Euler--Lagrange equations (4), and let $\CC_t$
be the surface $x=x(s,t)$. 
Let us apply formula (2) with the boundary surface $\CC_{t_1}\cup\CC_{t_2}$ and with 
$$
\begin{aligned}{}
(x^j(s,t,\ &\e), \varphi^i(x(s,t,\e),\e),\pi_i(x(s,t,\e),\e))\\
&=P(\e)(x(s,t),\varphi(s,t),\pi(s,t)),\\ 
\delta x^j(s)&=\delta\widetilde x^j_\alpha(x(s,t_\sigma),\varphi(s,t_\sigma)),\\
\delta\varphi^i(s)&=\delta\widetilde\varphi^i_\alpha(x(s,t_\sigma),\varphi(s,t_\sigma)),\ \ \sigma=1,2.
\end{aligned}
$$
Since the functional $J$ is preserved by the transformation $P(\e)$, we have $\delta J=0$. This implies that the functional
\begin{equation}
\begin{aligned}{}
\Phi_\alpha&=\Phi_\alpha(x^j(\cdot);\varphi^i(\cdot),\pi_i(\cdot))\\
&=\int_\CC\left(\sum\pi_i(s)a^i_\alpha(x(s),\varphi(s))-\sum H_j(s)b^j_\alpha(x(s),\varphi(s))\right)ds
\end{aligned}
\end{equation}
satisfies the equality 
$$
\Phi_\alpha(x^j(\cdot,t_1);\varphi^i(\cdot,t_1),\pi_i(\cdot,t_1))=\Phi_\alpha(x^j(\cdot,t_2);\varphi^i(\cdot,t_2),\pi_i(\cdot,t_2)),
$$
i. e. it satisfies equation (20). In other words, $\Phi_\alpha$ is an observable.

This observable generates the symmetry Hamiltonian flow, in the sense that
\begin{equation}
\left\{
\begin{array}{l}
\left.\frac{\partial\varphi^i(s,t,\e)}{\partial\e}\right|_{\e=0}=\frac{\delta\Phi_\alpha}{\delta\pi_i(s)}(x^j(\cdot,t);\varphi^i(\cdot,t),\pi_i(\cdot,t)),\\
\left.\frac{\partial\pi_i(s,t,\e)}{\partial\e}\right|_{\e=0}=-\frac{\delta\Phi_\alpha}{\delta\varphi^i(s)}(x^j(\cdot,t);\varphi^i(\cdot,t),\pi_i(\cdot,t)).
\end{array}
\right.
\end{equation} 
Equivalently, for an arbitrary observable $\Phi(x^j(\cdot);\varphi^i(\cdot),\pi_i(\cdot))$ and for any $P\in G$, the functional 
\begin{equation}
P\Phi(x^j(\cdot);\varphi^i(\cdot),\pi_i(\cdot))=\Phi(x^j(\cdot);\widetilde\varphi^i(\cdot),\widetilde\pi_i(\cdot))
\end{equation}
is also an observable, where 
\begin{equation}
(\widetilde x^j(\cdot),\widetilde\varphi^i(\cdot),\widetilde\pi_i(\cdot))=P(x(\cdot),\varphi(\cdot),\pi(\cdot)),
\end{equation}
and one has
\begin{equation}
\left.\frac{\partial P(\e)\Phi(x^j(\cdot,t);\varphi^i(\cdot,t),\pi_i(\cdot,t))}{\partial\e}\right|_{\e=0}=\{\Phi_\alpha,\Phi\}(x^j(\cdot,t);\varphi^i(\cdot,t),\pi_i(\cdot,t)).
\end{equation}

For $\alpha,\beta\in\g$, it is not difficult to check that
\begin{equation}
\{\Phi_\alpha,\Phi_\beta\}=\Phi_{[\alpha,\beta]}.
\end{equation}
\medskip

{\bf Examples.} 1) For the scalar field with self-action (10, 12, 25) such that $V(x,\varphi)=V(\varphi)$ is independent of $x$, the 
symmetry group $G$ is the nonhomogeneous Lorentz group $O(n,1)\widetilde\times\R^{n+1}$.

2) For the Plateau problem (16), the symmetry group is the group $O(N)\widetilde\times\R^N$ of isometries of the Euclidean space $\R^N$.

\section{Mathematical quantization of fields}

In this Section we construct mathematical quantization of classical field theories.

\setcounter{subsection}{-1}

\subsection{Differential calculus for functionals} Let $\Psi(\varphi(\cdot))$ be a functional depending on a smooth function $\varphi(s)$,
$s=(s^1,\ldots,s^n)$. We shall consider functions $\varphi(s)$ from a topological vector space $V$ of functions which is a nuclear space or a union of nuclear spaces [12], for example, the Schwartz space of 
smooth functions rapidly decreasing at infinity or the space of smooth functions
with compact support. We shall call the elements of the space $V$ by main functions, and the elements of the dual space $V'$ by distributions. One has the inclusion $V\subset V'$.

 The functional $\Psi$ is called differentiable if for an arbitrary main function $\varphi(s,\e)$ smoothly depending on a small 
parameter $\e$, one has
\begin{equation}
\delta\Psi(\varphi(\cdot))=\int\frac{\delta\Psi}{\delta\varphi(s)}(\varphi(\cdot))\delta\varphi(s)ds
\end{equation}
for some distribution $\delta\Psi/\delta\varphi(s)$ call\-ed the functional or variational derivative of $\Psi$ at the point $\varphi(\cdot)$, where 
$\delta\Psi$ and $\delta\varphi(s)$ denote the differential of $\Psi(\varphi(\cdot,\e))$ and $\varphi(s,\e)$ with respect to $\e$ at $\e=0$.

The functional $\Psi$ is called infinitely differentiable if for any $k$ one has
\begin{equation}
\delta^k\Psi(\varphi(\cdot))=\int\frac{\delta^k\Psi}{\delta\varphi(s_1)\ldots\delta\varphi(s_k)}(\varphi(\cdot))\delta\varphi(s_1)\ldots\delta\varphi(s_k)ds_1\ldots ds_k
\end{equation}
for some symmetric distribution $\delta^k\Psi/\delta\varphi(s_1)\ldots\delta\varphi(s_k)$ of $s_1,\ldots,s_k$ call\-ed the $k$-th functional or variational derivative of $\Psi$ at the point $\varphi(\cdot)$, where $\delta^k\Psi$
denotes the $k$-th differential of  $\Psi(\varphi(\cdot,\e))$ with respect to $\e$ at $\e=0$.

\subsection{Renormalization map and noncommutative multiplication of field theory Hamiltonians} The idea of this Subsection 
has been proposed in the papers [6--8].

Let $V$ be the space of smooth main functions $(\varphi^i(s)$, $\pi_i(s)$, $i=1,\ldots,m)$. Denote $\pi_i(s)=\rho_i(s)$, $\varphi^i(s)=\rho_{m+i}(s)$, $i=1,\ldots,m$. Let 
$H=H(\rho(\cdot))$ be an infinitely differentiable functional on the space $V$, where $\rho(s)=(\rho_i(s)$, $i=1,\ldots,2m)$.
We shall call $H$ a field theory {\it Hamiltonian}. A Hamiltonian $H$ is called {\it regular} if for any $k$ and for any $\rho(s)=(\rho_i(s))\in V$, the $k$-th functional derivatives
$\frac{\delta^k H}{\delta\rho_{i_1}(s_1)\ldots\delta\rho_{i_k}(s_k)}(\rho(\cdot))$ are main functions of $(s_1,\ldots,s_k)$. A Hamiltonian $H$ is called {\it classical} if it generates a well defined 
Hamiltonian flow 
\begin{equation}
\left\{\begin{array}{l}
\frac{\partial\varphi^i(s,t)}{\partial t}=\frac{\delta H}{\delta\pi_i(s)}(\varphi^i(\cdot,t),\pi_i(\cdot,t)),\\
\frac{\partial\pi_i(s,t)}{\partial t}=-\frac{\delta H}{\delta\varphi^i(s)}(\varphi^i(\cdot,t),\pi_i(\cdot,t))
\end{array}
\right.
\end{equation}
on the space $V$, i. e. if  $\frac{\delta H}{\delta\rho_i(s)}(\rho(\cdot))$ is a main function of $s$ smoothly depending on $\rho(\cdot)$. In other words, denote by 
$$
SV'=\bigoplus\limits_{k=0}^\infty S^k V',\ \ SV=\bigoplus_{k=0}^\infty S^k V 
$$
the topological symmetric algebra of the space $V'$ and respectively of $V$. Then a Hamiltonian $H$ is regular 
if and only if for any $k$ and any $\rho(\cdot)$ one has $\frac{\delta^k H}{\delta\rho_{i_1}(s_1)\ldots\delta\rho_{i_k}(s_k)}(\rho(\cdot))\in S^k V$, and $H$ is classical if and only if
for any $k$ and any $\rho(\cdot)$ one has
$$
\frac{\delta^k H}{\delta\rho_{i_1}(s_1)\ldots\delta\rho_{i_k}(s_k)}(\rho(\cdot))\in  S^k V'\cap(S^{k-1}V'\otimes V).
$$
An example of classical Hamiltonian is the Hamiltonian $H(t)$ given by (18) (for any $t$). 

Classical Hamiltonians form a Poisson algebra with respect to multiplication of 
Hamiltonians and the Poisson bracket (22),
\begin{equation}
\{H_1,H_2\}=\int\sum\limits_{i,j} \omega_{ij}\frac{\delta H_1}{\delta\rho_i(s)}\frac{\delta H_2}{\delta\rho_j(s)}ds,
\end{equation}
where $\omega_{ij}=\delta_{i,j-m}-\delta_{i-m,j}$. This Poisson algebra appeared in a non-rigorous form in the book [11], and in 
the rigorous form in [6--8]. Hamiltonian regularization, i.~e. approximation of field theory Hamiltonians by regular Hamiltonians, appeared in the book [14].

Define the {\it Weyl--Moyal algebra} as the associative algebra $SV$ of regular polynomial Hamiltonians with the Moyal product
\begin{equation}
\begin{aligned}{}
H_1 *_{Moyal}H_2(\rho(\cdot))=&\exp\left(-\frac{ih}2\int\sum\limits_{i,j} \omega_{ij}\frac{\delta}{\delta\rho_i^{(1)}(s)}\frac{\delta}{\delta\rho_j^{(2)}(s)}ds\right)\\
&\left.H_1(\rho^{(1)}(\cdot))H_2(\rho^{(2)}(\cdot))\right|_{\rho^{(1)}(\cdot)=\rho^{(2)}(\cdot)=\rho(\cdot)}.
\end{aligned}
\end{equation}
Here $ih=$(imaginary unit)$\times$(the Planck constant (a small parameter)).

Let 
\begin{equation}
G_N= G_N^{i_1,\ldots,i_N}(h,s_1,\ldots,s_N)=\sum\limits_{k=1}^\infty h^k G_{N,k}^{i_1,\ldots,i_N}(s_1,\ldots,s_N)\in S^N V',
\end{equation}
for $N=2,3,\ldots$, be real symmetric distributions depending on $h$. For a regular polynomial Hamiltonian
$H$, put 
\begin{equation}
\begin{aligned}{}
\widetilde H=\exp&\left(\sum\limits_{N=2}^\infty\frac1{N!}\int\sum\limits_{i_1,\ldots,i_N}G_N^{i_1,\ldots,i_N}(h,s_1,\ldots,s_N)\right.\\
&\left.\times\frac{\delta^N}{\delta\rho_{i_1}(s_1)\ldots\delta\rho_{i_N}(s_N)}ds_1\ldots ds_N\right)H.
\end{aligned}
\end{equation}
The Hamiltonian $\widetilde H$ is called the {\it renormalized} (regular) {\it Hamiltonian}, and $\widetilde H-H$ is called the {\it counterterm} (regular) {\it Hamiltonian}. The transform
$H\to\widetilde H$ (39) is called the {\it renormalization map}.

Let $H\to\widehat H$ be the inverse transform to the transform $H\to\widetilde H$, i.e. $\widehat H$ is given by formula (39) with $G_N$ replaced by $-G_N$.

Define the {\it $*$-product} of regular polynomial Hamiltonians by the formula
\begin{equation}
H_1*H_2=(\widehat H_1*_{Moyal}\widehat H_2)\widetilde{\ \ }.
\end{equation}
This is an associative product on the space $SV$ of regular polynomial Hamiltonians.

Explicitly, one has
\begin{equation}
\begin{aligned}{}
&H_1*H_2=\exp\left(-\frac{ih}2\int\sum\limits_{i,j} \omega_{ij}\frac{\delta}{\delta\rho_i^{(1)}(s)}\frac{\delta}{\delta\rho_j^{(2)}(s)}ds\right.\\ 
&+\left.\sum\limits_{N=2}^\infty\sum_{l=1}^{N-1}G_N^{(l,N-l)}\right)\left.H_1(\rho^{(1)}(\cdot))H_2(\rho^{(2)}(\cdot))\right|_{\rho^{(1)}(\cdot)=\rho^{(2)}(\cdot)=\rho(\cdot)},
\end{aligned}
\end{equation}
where 
\begin{equation}
\begin{aligned}{}
&G_N^{(l,N-l)}=\frac1{l!(N-l)!}\sum\limits_{i_1,\ldots,i_N}\int G_N^{i_1,\ldots,i_N}(h,s_1,\ldots,s_N)\\
&\times\frac{\delta^l}{\delta\rho^{(1)}_{i_1}(s_1)\ldots\delta\rho^{(1)}_{i_l}(s_l)}
\frac{\delta^{N-l}}{\delta\rho^{(2)}_{i_{l+1}}(s_{l+1})\ldots\delta\rho^{(2)}_{i_N}(s_N)}ds_1\ldots ds_N.
\end{aligned}
\end{equation}

One has
\begin{equation}
[H_1,H_2]\eqdef H_1*H_2-H_2*H_1=-ih\{H_1,H_2\}+o(h).
\end{equation}

{\bf Definition.} The transform (39) is called a {\it finite renormalization} if all the distributions $G_N$ are main functions. Two renormalizations 
corresponding to distributions $G_N'$ and $G_N''$ are called {\it equivalent} if they differ by a finite renormalization (39) with $G_N=G_N'-G_N''$.
Finite renormalizations form a commutative group called the {\it renormalization group}.

\subsection{Quantization of classical field theory equations}
Consider a classical field theory with the action functional (1) and with the generalized Hamilton--Jacobi equation (9).
Let (40) be the $*$-product corresponding to a renormalization (39). Let us call by the {\it renormalized Tomonaga--Schwinger equation} the system of equations
\begin{equation}
\left\{
\begin{array}{l}
ih\sum\limits_j
x^j_{s^k}\frac{\delta\Phi}{\delta x^j(s)}=\sum\limits_i\varphi^i_{s^k}\pi_i(s)*\Phi+c_k(s,h)\Phi,\ \ \ k=1,\ldots,n,\\
\HH\left(x^j(s),\varphi^i(s),x^j_{s^k}(s),\varphi^i_{s^k}(s),
\pi_i(s),-ih\frac{\delta}{\delta x^j(s)}\right)*\Phi\\
+C(s,x^j(s),\varphi^i(s),\pi_i(s),x^j_{s^k}(s),\varphi^i_{s^k}(s),\pi_{is^k}(s),\ldots,h)*\Phi=0
\end{array}
\right.
\end{equation}
for a functional $\Phi(x^j(\cdot);\varphi^i(\cdot),\pi_i(\cdot))$, where 
\begin{equation}
C(\ldots,h)=C(s,x^j(\cdot),\varphi^i(\cdot),\pi_i(\cdot),h)
\end{equation}
 is a function 
of $s$, $x^j(s)$, $\varphi^i(s)$, $\pi_i(s)$ and their partial derivatives of finite orders at the point $s$ with $C(\ldots,0)=0$, and 
$c_k(s,h)$, $k=1,\ldots,n$,  are distributions in $s$ with $c_k(s,0)=0$.
We assume, without loss of generality,\footnote{because the $(n+1)$-th equation $\HH=0$ in the generalized Hamilton--Jacobi equation is defined not uniquely but only modulo the first $n$ equations.} that 
\begin{equation}
\HH=\HH(\ldots,-H_j(s))=\sum_{j=0}^n C_j(s)H_j(s)+A(s),
\end{equation}
where $C_j(s)$ and $A(s)$ do not depend on $H_{j'}(s)$, and for any $j$ the coefficient $C_j(s)$ before $H_j(s)$ does not depend on $x^j(\cdot)$.
For the scalar field with self-action (10--15) this condition is satisfied. 
\medskip

{\bf Agreement.} The system of equations (44), as well as the subsequent constructions and formulas, are understood as the formal limit of regularized formulas,
with the Hamiltonians $H(\rho(\cdot))$ replaced by regular Hamiltonians $H^\Lambda_{reg}(\rho(\cdot))$, $\Lambda>0$ such that $\lim\limits_{\Lambda\to0}H^\Lambda_{reg}=H$.
We say that a formal formula including irregular Hamiltonians is {\it well defined} if the limit as $\Lambda\to0$ of the regularized formula exists and does not depend on the 
choice of regularization.  
\medskip

The system of equations (44) is, in general, equivalent to the system
\begin{equation}
ih\frac{\delta\Phi}{\delta x^j(s)}=H_j(s,h)*\Phi,
\end{equation}
where $H_j(s,h)$ is a deformation of the function (17), i.~e. to the equation
\begin{equation}
ih\frac{\partial\Phi(x(\cdot,t),\varphi^i(\cdot),\pi_i(\cdot))}{\partial t}=H(t,h)*\Phi
\end{equation}
for any function $x(s,t)$, where $H(t,h)$ is a deformation of the Hamiltonian (18).

{\bf Example.} For the scalar field (10, 12), the renormalized Tomonaga--Schwinger equation is 
\begin{equation}
\left\{
\begin{array}{l}
ihx^\mu_{s^k}(s)\frac{\delta\Phi}{\delta x^\mu(s)}=
\varphi_{s^k}(s)\pi(s)*\Phi+c_k(h,s)\Phi,\ \ k=1,\ldots,n,\\
ih\vol\cdot\frac{\delta\Phi}{\delta\n(s)}=
\left(\frac12\pi(s)^2
+ \vol^2\left(\frac12 d\varphi(s)^2+V(x(s),\varphi(s))\right)\right)*\Phi.
\end{array}
\right.
\end{equation}

Let $\Phi(x(\cdot);\rho(\cdot))=U(x_0(\cdot),x(\cdot);\rho(\cdot))$
be the (formal) evolution operator for equation (48),
with the initial condition 
\begin{equation}
U(x_0(\cdot),x_0(\cdot);\rho(\cdot))=1.
\end{equation}

Explicitly, for two surfaces $x_0(s)=x(s,t_0)$ and $x(s)=x(s,t_1)$, one has
\begin{equation}
\begin{aligned}{}
&U(x_0(\cdot),x(\cdot))=T\exp*\int\limits_{t_0}^{t_1} \frac1{ih}H(\tau,h)d\tau\\
&\eqdef\sum\limits_{k=0}^\infty\frac1{(ih)^k}\int\limits_{t_0\le \tau_1\le\ldots\le\tau_k\le t_1}H(\tau_k,h)*\ldots*H(\tau_1,h)\,d\tau_1\ldots d\tau_k.
\end{aligned}
\end{equation}
\medskip

{\bf Definition.} The renormalized Tomonaga--Schwinger equation (44) is called {\it integrable} if the operator $U(x_0(\cdot)$, $x(\cdot);$ $\rho(\cdot))$ is correctly defined up to multiplication by a constant 
(depending on $x_0(\cdot)$ and $x(\cdot)$), 
i.~e. if it depends, up to a constant, not on the function $x(s,t)$ but only on the
initial surface $x=x_0(s)$ and the final surface $x=x(s)$. 
\medskip

{\bf Theorem 1.} {\it For any renormalization \emph{(39)}, the renormalized To\-mo\-na\-ga--Schwinger equation for the scalar field \emph{(49)} is integrable.}

{\it Proof.} The system of equations (47) is equivalent to the system
\begin{equation}
ih\frac{\delta\widehat\Phi}{\delta x^j(s)}=\widehat H_j(s)*_{Moyal}\widehat\Phi.
\end{equation}
We must check the Frobenius integrability (constant curvature) condition
\begin{equation}
ih\frac{\delta\widehat H_j(s)}{\delta x^{j'}(s')}-ih\frac{\delta\widehat H_{j'}(s')}{\delta x^j(s)}+[\widehat H_j(s),\widehat H_{j'}(s')]_{Moyal}=const.
\end{equation} 
For the scalar field with self-action, $\widehat H_j(s)$ is the sum of a quadratic expression in $\varphi(\cdot)$, $\pi(\cdot)$, a constant, and a function of $\widehat V(x(s),\varphi(s))$. This implies that 
\begin{equation}
[\widehat H_j(s),\widehat H_{j'}(s')]_{Moyal}=-ih\{\widehat H_j(s),\widehat H_{j'}(s')\}, 
\end{equation}
because the terms with $h^3$, $h^5$, etc. in the Moyal commutator vanish. Hence condition (53) follows from the classical Frobenius 
integrability condition (24) with $V(x(s),\varphi(s))$ replaced by $\widehat V(x(s),\varphi(s))$. Q.E.D.
\medskip

The function $U(x_0(\cdot),x(\cdot);\rho(\cdot))$ also satisfies the equations
\begin{equation}
ih\frac{\delta U(x_0(\cdot),x(\cdot))}{\delta x^j_0(s)}=-U(x_0(\cdot),x(\cdot))*H_j(s,h,x_0(\cdot)),
\end{equation}
\begin{equation}
U(x_0(\cdot),x_2(\cdot))=U(x_1(\cdot),x_2(\cdot))*U(x_0(\cdot),x_1(\cdot)).
\end{equation}

The function $U(x_0(\cdot),x(\cdot);\rho(\cdot))$ is covariant with respect to the left and right actions of a central extension of the Lie algebra
of smooth vector fields
\begin{equation}
\int\sum\limits_{i,k}a_k(s)\varphi^i_{s^k}(s)\pi_i(s)ds,
\end{equation}
with main functions $a_k(s)$, $k=1,\ldots,n$, on the space of functions $\Phi(\rho(\cdot))$.
\medskip

{\bf Problem.} For the scalar field, formula (51) for $U(x_0(\cdot),x(\cdot);\rho(\cdot))$ makes sense only for space-like surfaces $x(s,\tau)$ for any $\tau$, because for non-space-like surfaces the 
Hamiltonian $H(\tau)$ is not well defined. However, equations (49) given by formulas (13--15) make sense for an arbitrary surface $x(s)$. 
Can the operator $U(x_0(\cdot),x(\cdot);\rho(\cdot))$ be extended to a (formal) solution of equations (47--50, 55, 56) for non-space-like surfaces $x_0(s)$, $x(s)$?
\medskip

{\bf Definition.} A {\it quantization} of a classical field theory given by the variational principle (1) is an integrable renormalized To\-mo\-na\-ga--Schwi\-nger equation (44).  
Two quantizations are called {\it equivalent} if they differ by a finite renormalization.
\medskip

 Let us call by a (formal) quantum field theory {\it observable} a functional $\Phi(x(\cdot);$ $\rho(\cdot))$ satisfying the {\it generalized Heisenberg equation}
\begin{equation}
ih\frac{\delta\Phi}{\delta x^j(s)}=[H_j(s,h),\Phi].
\end{equation}
For these observables, one has
\begin{equation}
\Phi(x(\cdot))=U(x_0(\cdot),x(\cdot))*\Phi(x_0(\cdot))*U(x_0(\cdot),x(\cdot))^{-1}.
\end{equation}
Observables form an associative algebra with respect to the $*$-product. For any surface $x_0(s)$, this algebra is identified with the algebra of Hamiltonians 
$H(\rho(\cdot))=\Phi(x_0(\cdot);\rho(\cdot))$. For equivalent quantizations of a classical field theory, the algebras of observables 
are naturally identified.

Define the {\it conjugate observable} $\Phi^*$ to an observable $\Phi$ as the complex conjugate functional. One has
\begin{equation}
(\Phi_1*\Phi_2)^*=\Phi_2^* *\Phi_1^*.
\end{equation} 

Any quantum field theory observable $\Phi$ is covariant with respect to the natural action of the group of diffeomorphisms of the variables $(s^1,\ldots,s^n)$, with the Lie algebra 
consisting of smooth vector fields (57).

\subsection{Quantization on surfaces of constant time} 
Denote $x^0=t$, $(x^1,\ldots,x^n)=\x$. Denote by $U(t_0,t)$ 
the evolution operator for flat surfaces of constant time,
\begin{equation}
U(t_0,t)=U(x_0(\cdot),x(\cdot)),\ \ x_0(s)=(t_0,s),\ \ x(s)=(t,s).
\end{equation}
The function $U(t_0,t)=\Phi(t)$ is the evolution operator for the Schr\"odinger equation
\begin{equation}
ihd\Phi/dt=H(t,h)*\Phi,
\end{equation}
where $H(t,h)=H(t,\varphi(\cdot),\pi(\cdot),h)$ is the Hamiltonian for the surface $x(s)=(t,s)$ of constant time.

One has
\begin{equation}
U(t_0,t)=T\exp*\int\limits_{t_0}^t\frac1{ih} H(\tau,h)d\tau.
\end{equation}

{\bf Example.} For the scalar field (10, 49), one has 
\begin{equation}
H(t,h)=\int\left(\frac12\pi(\x)^2+\frac12\sum\limits_{j=1}^n\varphi_{x^j}(\x)^2+V(t,\x,\varphi(\x))\right)d\x.
\end{equation}

Any quantum field theory observable $\Phi(x(\cdot);\rho(\cdot))$ is uniquely determined by its values for flat surfaces of constant time,
\begin{equation}
\Phi(t_0,\rho(\cdot))=\Phi(x(\cdot);\rho(\cdot)),\ \ x(s)=(t_0,s).
\end{equation}
The function $\Phi(t_0,\rho(\cdot))$ satisfies the Heisenberg equation
\begin{equation}
ihd\Phi/dt_0=[H(t_0,h),\Phi].
\end{equation}
One has
\begin{equation}
\Phi(t)=U(t_0,t)*\Phi(t_0)*U(t_0,t)^{-1}.
\end{equation} 

\section{Mathematical expressions for the Bogoliubov $S$-matrix, the Wightman functions, and the Green functions}

\subsection{The half $S$-matrix}

Consider the $\varphi^d$ model of scalar field in $\R^{n+1}$ with variable interaction, i.~e. the  Lagrangian $F(x^j,\varphi,\varphi_{x^j})$ (10) with 
\begin{equation}
V(x,\varphi)=\frac{m^2}2\varphi^2+\frac{g(x)}{d!}\varphi^d,
\end{equation}
where $g(x)$ is a real smooth function with compact support called the interaction cutoff function. 

Let
\begin{equation}
H(t,\varphi(\cdot),\pi(\cdot))=H_0(\varphi(\cdot),\pi(\cdot))+\int\frac{g(t,\x)}{d!}\varphi(\x)^d d\x
\end{equation}
be the Hamiltonian (64) of scalar field, where 
\begin{equation}
H_0(\varphi(\cdot),\pi(\cdot))=\int\left(\frac12\pi(\x)^2+\frac12\sum\limits_{j=1}^n\varphi_{x^j}(\x)^2+\frac{m^2}2\varphi(\x)^2\right)d\x
\end{equation}
is the Hamiltonian of free scalar field.

Consider an arbitrary renormalization (39) and the quantized scalar field theory (49). Let $\Phi(t)$ be an observable for this quantum field theory. 
For $t=t_0<<0$, $\Phi(t_0)$ is identified with an observable $\Phi_0(t_0)$ of the free quantum field theory (i.~e. the theory with $g(x)\equiv0$). For $t=t_1>>0$,
$\Phi(t_1)$ is identified with another observable $\Phi_1(t_1)$ of the free quantum field theory. The observables $\Phi_0(t)$ and $\Phi_1(t)$ are related by the equality
\begin{equation}
\begin{aligned}{}
\Phi_1(t_0)&=U(t_0,t_1)^{-1}*\Phi_0(t_1)*U(t_0,t_1)\\
&=S^{1/2}(t_0,t_1)^{-1}*\Phi_0(t_0)*S^{1/2}(t_0,t_1),
\end{aligned}
\end{equation}
where $U(t_0,t)$ is the evolution operator (63) for scalar quantum field theory, 
\begin{equation}
S^{1/2}(t_0,t)=U_0(t_0,t)^{-1}*U(t_0,t),
\end{equation}
and 
\begin{equation}
U_0(t_0,t)=\exp*\frac{t-t_0}{ih}H_0
\end{equation}
is the evolution operator for free quantum field theory. The functional $S^{1/2}(t_0,t)=\Phi(t_0)$ is a (formal) free quantum field theory observable depending on $t$ called the {\it half $S$-matrix}.

\subsection{Lorentz covariance property.} It is the following assumption.

{\it The quantization \emph{(49)} of the free classical field theory \emph{(10, 12, 25, 68)} with $g(x)\equiv0$, given by renormalization \emph{(39)}, is Lorentz covariant, i.~e.
for any observable $\Phi(x^j(\cdot);$ $\varphi(\cdot),$ $\pi(\cdot))$ of free quantum field theory and for any element $P$ of the nonhomogeneous 
Lorentz symmetry group $G=O(n,1)\widetilde\times\R^{n+1}$, one has
\begin{equation}
P\Phi(x^j(\cdot);\varphi(\cdot),\pi(\cdot))=\Phi(P^{-1}x(\cdot);\varphi(\cdot),\pi(\cdot)),
\end{equation}
where $P\Phi$ is given by \emph{(29)}. }

{\bf Corollary.} {\it For any element $\alpha$ of the Lie algebra $\g$ of the group $G$ and for any functional $\Phi(\varphi(\cdot),\pi(\cdot))$, one has
\begin{equation}
[\Phi_\alpha(t_0),\Phi]=-ih\{\Phi_\alpha(t_0),\Phi\},
\end{equation} 
where $\Phi_\alpha(t_0)$ is the generator \emph{(27)} of the symmetry $\alpha$. In particular, for the Hamiltonian
$H_0$ \emph{(70)}, one has}
\begin{equation}
[H_0,\Phi]=-ih\{H_0,\Phi\}.
\end{equation}

This Corollary implies that any free quantum field theory observable $\Phi$ can be considered as a functional of a solution $\varphi_0(x)$ of the Euler--Lagrange equation with $g(x)\equiv0$,
\begin{equation}
\frac{\partial^2\varphi_0}{\partial t^2}-\sum\limits_{j=1}^n\frac{\partial^2\varphi_0}{\partial x^{j2}}+m^2\varphi_0=0,
\end{equation}
 i.~e. the Klein--Gordon equation,
with the initial conditions
\begin{equation}
\varphi_0(t_0,\x)=\varphi(\x),\ \ \partial\varphi_0/\partial t(t_0,\x)=\pi(\x).
\end{equation}
More concretely, consider the Schwartz space of functions $(\varphi(\x),\pi(\x))$. By (78) this space is identified with the space $V$ of smooth solutions $\varphi_0(x)$ of the Klein--Gordon equation (77)
rapidly decreasing in the space directions. The free quantum field theory observables  $\Phi(t_0,\varphi(\cdot),\pi(\cdot))=\Phi(\varphi_0(x))$  are polynomial Hamiltonians on $V$. 
The space of observables $\Phi$ has a family of product operations depending on the Planck constant $h$. 
For $h=0$ it is the usual commutative product, and for $h\ne0$ it is the noncommutative $*$-product.

The Lorentz covariance property means that the distributions $G_N\in S^N V'$, $N=2,3,\ldots,$ are Lorentz invariant.

\subsection{The Bogoliubov $S$-matrix, the Wightman and Green functions}
The half $S$-matrix $S^{1/2}(t_0,t)$ (72)
is a (formal) free quantum field theory observable given by the formula
\begin{equation}
S^{1/2}=S^{1/2}[H_{int}](t)=T\exp*\int\limits_{-\infty}^t\frac1{ih}H_{int}(\tau)d\tau,
\end{equation}
where
\begin{equation}
\begin{aligned}{}
H_{int}(t)&=U_0(t_0,t)^{-1}*\int\frac{g(t,\x)}{d!}\varphi(\x)^d d\x*U_0(t_0,t)\\
&=\int\frac{g(t,\x)}{d!}\varphi_0(t,\x)^d d\x
\end{aligned}
\end{equation}
is the interaction Hamiltonian.

 Let $\F$ be the Fock Hilbert space for free scalar field with mass $m$ in the space-time $\R^{n+1}$. Normally ordered linear operators in $\F$ can be identified with their Wick symbols, 
which are functionals $\Phi(\varphi_0(\cdot))$ of a solution $\varphi_0(x)$ of the Klein--Gordon equation (77). The product of normally ordered operators in the space $\F$ corresponds to the product
$\Phi_1*_\F\Phi_2$ of their symbols $\Phi_1$, $\Phi_2$, given by formula (40) for the transform (39) with the trivial functions $G_3^\F= G_4^\F=\ldots=0$ and with a 
nontrivial function $G_2^\F$. Hence we have a $G$-invariant not everywhere defined (discontinuous) homomorphism $\Phi\to\widehat\Phi$ from the algebra 
of free quantum field theory observables corresponding to a Lorentz covariant renormalization $G_2$, $G_3$, $\ldots$ (39) 
to the algebra of Wick symbols of linear operators in $\F$.
The inverse homomorphism $\Phi\to\widetilde\Phi$ is given by formula (39) with the functions $G_2-G_2^\F$, $G_3$, $G_4$, $\ldots$. We shall call this homomorphism by renormalization.

Denote by 
\begin{equation}
\langle\Phi\rangle=\Phi(\varphi_0\equiv0)
\end{equation}
the vacuum average of an operator in $\F$ with the Wick symbol $\Phi$.

The linear operator in the Fock space
\begin{equation}
S[g(\cdot)]=\widehat{S^{1/2}}[\widetilde H_{int}](t_1)=\left(T\exp*\int\limits_{-\infty}^\infty\frac1{ih}\frac{g(t,\x)}{d!}\widetilde{\varphi_0(t,\x)^d}dt d\x\right)^{\widehat{\ \ \ }},
\end{equation}
where $t_1>>0$, is called the {\it Bogoliubov $S$-matrix} (with variable interaction) for the $\varphi^d$ model in $\R^{n+1}$.

Let $\jj(x)$ be a real smooth function with compact support called the source function. Put
\begin{equation}
\begin{aligned}{}
Z[H_{int},\jj(\cdot)]&=S^{1/2}[H_{int}](t_1)^{-1}*S^{1/2}\left[H_{int}+\int\jj(t,\x)\varphi_0(t,\x)d\x\right](t_1)\\
&=T\exp*\int\limits_{-\infty}^\infty\frac1{ih}\jj(t,\x)\varphi(t,\x)dtd\x,
\end{aligned}
\end{equation}
where $t_1>>0$ and
\begin{equation}
\begin{aligned}{}
\varphi(t,\x)&=U(t_0,t)^{-1}*\varphi(\x)*U(t_0,t)\\
&=S^{1/2}[H_{int}](t)^{-1}*\varphi_0(t,\x)*S^{1/2}[H_{int}](t).
\end{aligned}
\end{equation}

The distribution
\begin{equation}
\begin{aligned}{}
&W_N(x_1,\ldots,x_N)\\
&=\lim\limits_{g(x)\to g}\left\langle\left(\left.\frac{\delta Z[\widetilde H_{int},\jj(\cdot)]}{\delta\jj(x_1)}\right|_{\jj(x)\equiv 0}
*\ldots*\left.\frac{\delta Z[\widetilde H_{int},\jj(\cdot)]}{\delta\jj(x_N)}\right|_{\jj(x)\equiv 0}\right)^{\widehat{\ \ \ }}\right\rangle
\end{aligned}
\end{equation}
is called the {\it Wightman function}, and the distribution
\begin{equation}
V_N(x_1,\ldots,x_N)=\lim\limits_{g(x)\to g}\left\langle\left(\left.\frac{\delta^N Z[\widetilde H_{int},\jj(\cdot)]}{\delta\jj(x_1)\ldots\delta\jj(x_N)}\right|_{\jj(x)\equiv 0}\right)^{\widehat{\ \ \ }}\right\rangle
\end{equation}
is called the {\it Green function}.
\medskip

The following Theorem is obvious.
\medskip

{\bf Theorem 2.} {\it If the Lorentz covariance assumption holds and if the Bogoliubov 
$S$-matrix \emph{(82)}, the Wightman functions \emph{(85)}, and the Green functions \emph{(86)} are well defined, then they satisfy the 
axioms of quantum field theory \emph{[9, 10].}}
\medskip

Actually the purpose of perturbative quantum field theory is to construct a quantized field theory power series (38, 39, 49) such that the $S$-matrix (82), the Wightman functions (85), and the Green functions (86) are well defined
as power series, while axiomatic quantum field theory
studies the properties of these functions.
  
\subsection{Discussion}
The problem with quantities (82, 85, 86) is that they are, in general, ill-defined, because the renormalization $\Phi\to\widetilde\Phi$ and the inverse map $\Phi\to\widehat\Phi$ are discontinuous not everywhere defined maps. 
To obtain an expression for these quantities, one should choose a regularization $H^\Lambda_{int}$ of the interaction Hamiltonian $H_{int}=\lim\limits_{\Lambda\to0}H^\Lambda_{int}$. 
The limit as $\Lambda\to0$ of regularized expressions (82, 85, 86), in general, depends on the choice of regularization. This corresponds to the fact that the notions 
of vacuum and asymptotical non-interacting particles in quantum field theory are unnatural.\footnote{This argument is taken from [10].} The quantum field theory $S$-matrix describes scattering amplitudes of colliding particles. 
They are assumed to be non-interacting at large distances. However, 
the particles always interact with vacuum. In quantum field theory there is no natural notion of particles or vacuum but there is only the universal form of matter consisting of interacting quantum fields. Hence the definition 
of a quantum field theory from \S2 well agrees with the physical picture.  

The notion of vacuum is natural in quantum statistical mechanics. I heard [15] that the essential difference between quantum field theory and critical phenomena in quantum statistical mechanics has been 
verified experimentally.
\medskip

{\bf Conjecture.} {\it Axioms of quantum field theory can be, in general, neither mathematically realized nor disproved, like Continuum Hypothesis.}

This Conjecture means that ``final mathematical theory of everything'' does not exist.
\medskip

Mathematical quantum field theory in the Minkowski space-time is a hyperbolic theory. It studies the solutions $U(x^0(\cdot), x(\cdot))$ of the Cauchy problem and the related well defined quantities
(i.~e. quantities independent of regularization).
The following three questions are concerned with elliptic theories, with a different behavior. The natural problem for them is the boundary problem. 

{\bf Problem 1.} Construct and study mathematical quantization of field theory in the Euclidean space-time.

{\bf Problem 2.} Construct and study mathematical quantization of the Plateau problem (16).

{\bf Problem 3.} Construct and study mathematical quantization of the Dirichlet principle.

\end{document}